\documentclass[a4paper]{article}

\usepackage{icrc2013}
\usepackage{color}

\title{All Sky Cameras for the characterization of the Cherenkov Telescope Array candidate sites}

\shorttitle{All Sky Camera for CTA}

\authors{
Dusan Mandat$^{1}$,
Miroslav Pech$^{1}$,
Jan Ebr$^{1}$,
Miroslav Hrabovsky$^{1}$,
Michael Prouza$^{1}$,
Tomasz Bulik$^{2}$,
Ingomar Allekotte$^{3}$,
for the CTA Consortium.
}

\afiliations{
$^1$ Institute of Physics of Academy of Science of The Czech Republic, Czech Republic. \\
$^2$ Astronomical Observatory University of Warsaw, Poland. \\
$^3$ Centro At\'omico Bariloche, Comisi\'on Nacional de Energ\'ia At\'omica, San Carlos de Bariloche, Argentina. \\

}

\email{mandat@fzu.cz}

\abstract{The All Sky Camera (ASC) was developed as a universal device for the monitoring of the night sky quality. Eight ASCs are already installed and measure night sky parameters at eight of the candidate sites of the Cherenkov Telescope Array (CTA) gamma-ray observatory.  The ACS system consists of an astronomical CCD camera, a fish eye lens, a control computer and associated electronics. The measurement is carried out during astronomical night. The images are automatically taken every $5$ minutes and automatically processed using the control computer of the device. The analysis results are the cloud fraction (the percentage of the sky covered by clouds) and night sky brightness (in mag/arcsec$^{2}$). }

\keywords{All Sky Camera, CTA, candidate sites, cloud fraction.}

\begin{document}
\maketitle

\section{Cloudiness characterization of CTA candidate sites}

The future CTA observatory \cite {bib:cta} will include two observatories. One on the Southern and one on the Northern Hemisphere. There are nine candidate sites for the future CTA observatories - five at the Southern and four at the Northern Hemisphere. The goal of Site Selection Work Package \cite {bib:bulik} (SITE WP) is to characterize the candidate sites and produce relevant data for the Site Selection Committee. One of the important characteristic of the candidate site is the cloud fraction of the night sky and the amount of cloudless nights or more specifically cloudless time in a year. The ASC is an automated instrument for measurement of this parameter. Starting in November $2011$ the following sites are being characterized using ASCs :

\begin{table}[h]
\begin{center}
\begin{tabular}{|l|c|c|}
\hline Site & date of installation & GPS \\ \hline
Aar   & $11/2011$ &$26^{\circ}41.5'$S $16^{\circ}26.4'$E\\ \hline
SAC   & $02/2012$ &$24^{\circ}2.7'$S $66^{\circ}14.1'$W\\ \hline
SPM & $09/2012$ &$31^{\circ}0.9'$N $116^{\circ}31.3'$W\\ \hline
AZE & $09/2012$ &$35^{\circ}2.8'$N $111^{\circ}2.7'$W\\ \hline
AZW & $09/2012$ &$35^{\circ}8'$N $112^{\circ}51.9'$W\\ \hline
TEN & $10/2012$ &$28^{\circ}16.7'$N $16^{\circ}32'$W\\ \hline
CAS & $11/2012$ &$31^{\circ}42.2'$S $69^{\circ}15.5'$W\\ \hline
ARM & $04/2013$ &$24^{\circ}34.4'$S $70^{\circ}14.3'$W\\ \hline

\end{tabular}
\caption{Candidate sites for CTA. Aar - Aar farm, Namibia, SAC - San Antonio de Los Cobres, Argentina, SPM - SAN Pedro Martir, Mexico, AZE - Meteor Crater, USA, AZW - Yavapai ranch, USA,  TEN - Tenerife Spain, CAS - CASLEO observatory, Argentina, ARM - Cerro Armazones, Chile .}
\label{table_one}
\end{center}
\end{table}

\subsection{System Description}

\begin{figure}[t]
\begin{center}
\includegraphics[width=0.4\textwidth]{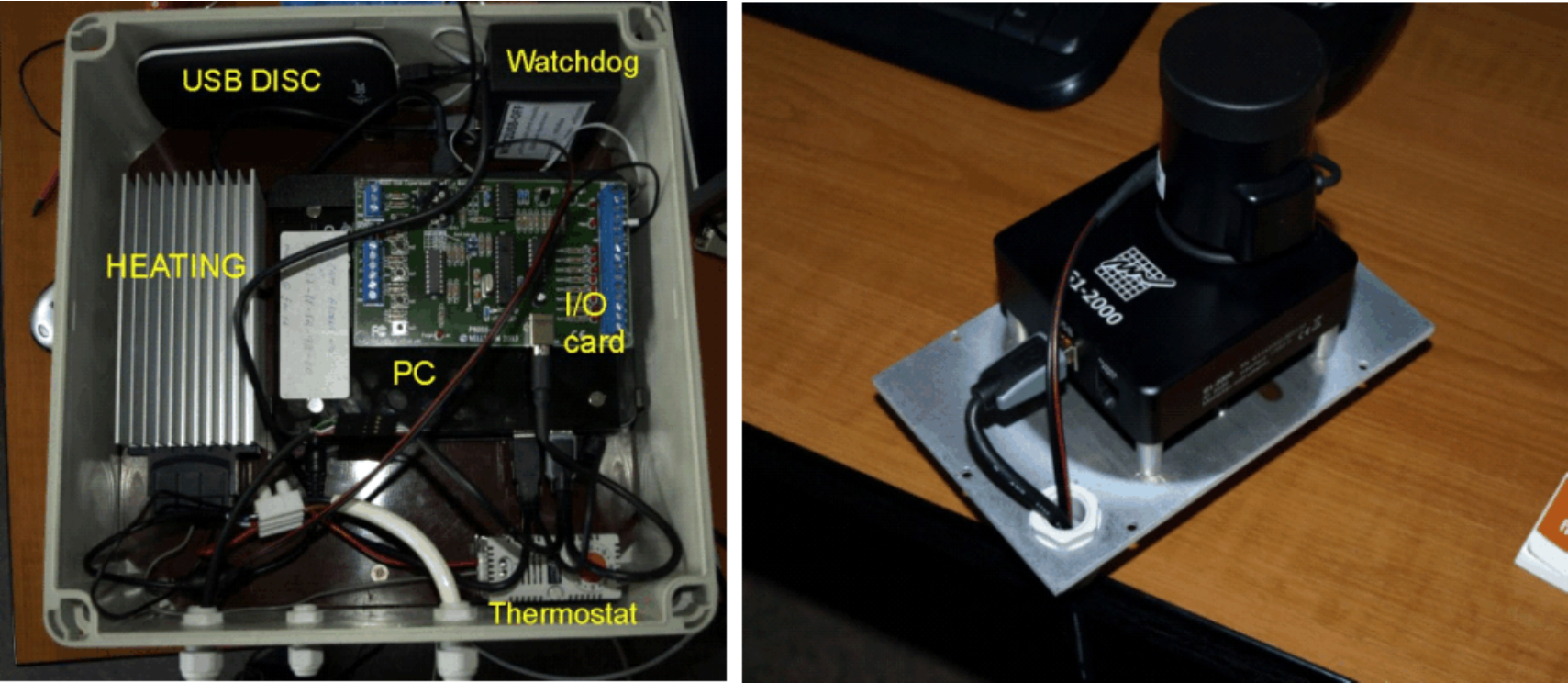}
\caption{Electronics and the inside of the camera box.}
\label{electronics}
\end{center}
\end{figure}

The basis of the system is an astronomical camera G$1-2000$ (Moravian Instruments a.s., Czech Republic, www.mii.cz), which is equipped with a CCD chip ICX$274$AL by SONY company. This chip has quantum efficiency higher than $50 \%$  at $450$--$550$ nm and its spectral sensitivy covers the range from approximately 400 to 900 nm. It has low read-out noise, a 16-bit ADC and a resolution of $1600\times1200$ pixels. The camera is equipped with a fish-eye varifocal lens (the field of view is $185$ degrees) and an electronic iris control.  This setup is capable of detecting a star with visual magnitude 6.3 in zenith. 

A  miniPC computer with USB $I/O$ controls the system and processes the data. The $I/O$ card controls the iris and the power switch of the camera. All electronics and a camera body (see Figure \ref{electronics}) are weatherproof. The system power is supplied by a solar power system.  The switching electronics turns the system ON after sunset and keeps it going during astronomical night. The system is OFF during the day (to save energy). Temperature of the electronics is controlled and stabilized using internal heating system during winter time to protect the ASC system. The camera is a part of an Atmoscope instrument. The Atmoscope measures local weather conditions and night sky brightness parameters. The Atmoscope is connected to central data storage using GPRS or WIFI connection depending on Atmoscope local position and nearby facilities. The quality of the link limits the amount and speed of the data transfer. 
\\

\subsection{Calibration of the camera}

The fish-eye varifocal lens has $185$ degrees field of view (FOV) and its aberration distorts the image. The correct transformation of the incident angle to the imaged pixel position is very important. The mechanism of the calibration is as follows: the light source (a spot as small as possible) is placed at a defined distance (at least $10$ m), then the body of camera is rotated around the axis such that the incident light angle (corresponding to the FOV angle) varies between $0$ and $90$ degrees. The position of the light spot is analyzed for each anglular step. This procedure is repeated for different plane cuts of the lens FOV. The final data are analyzed and fitted with a polynomial function. This calibration is regularly checked on-site by comparing the position of detected stars with the stars in a catalogue. On-site measurements are also used to determine the variation of the sensitivty of the system with the zenith angle.
\\

\subsection{Data acquisition and principles of analysis}

\begin{figure}[t]
\begin{center}
\includegraphics[width=0.45\textwidth]{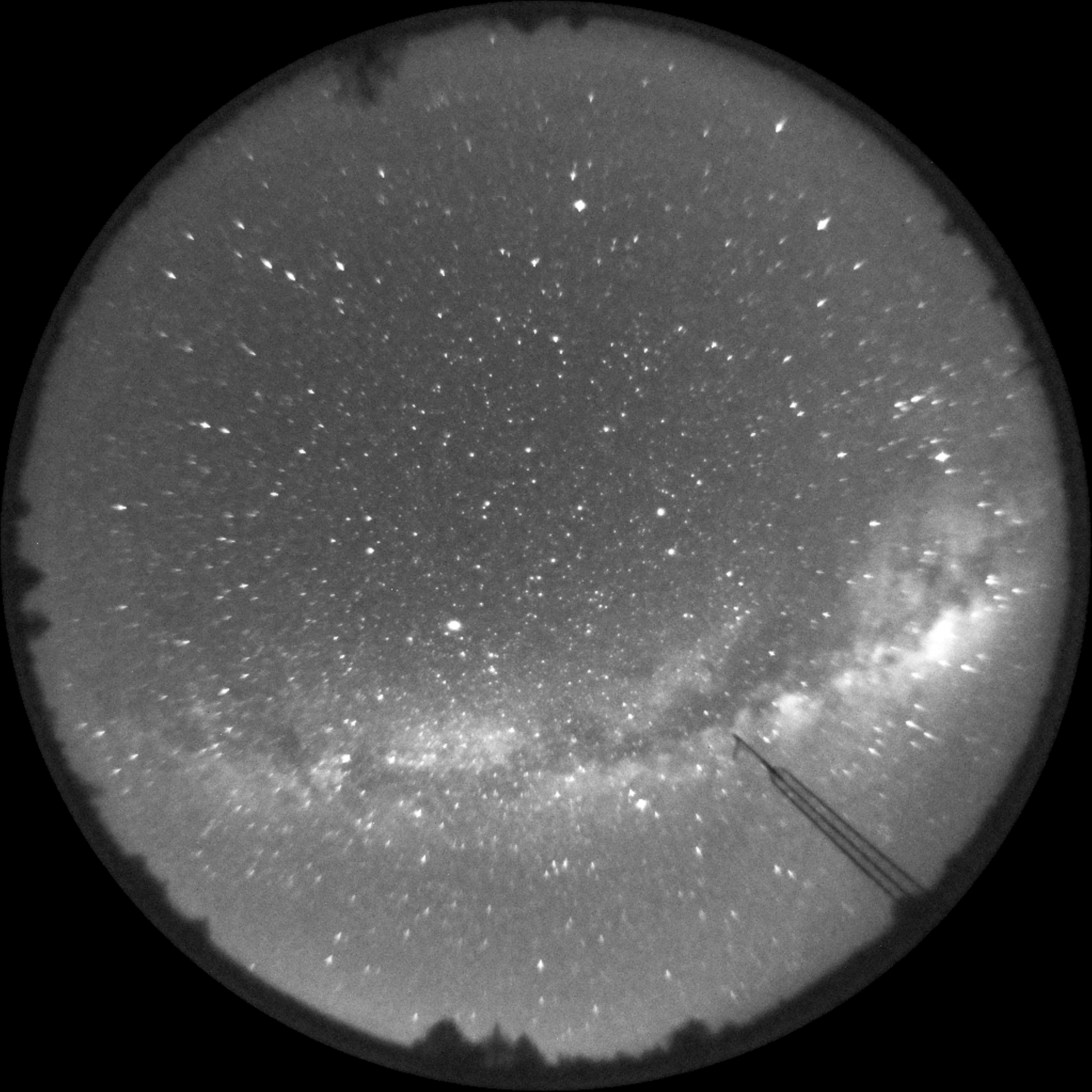}
\caption{Image of clear night sky from San Pedro Martir.}
\label{nightSky}
\end{center}
\end{figure}

\begin{figure}[!b]
\begin{center}
\includegraphics[width=0.46\textwidth]{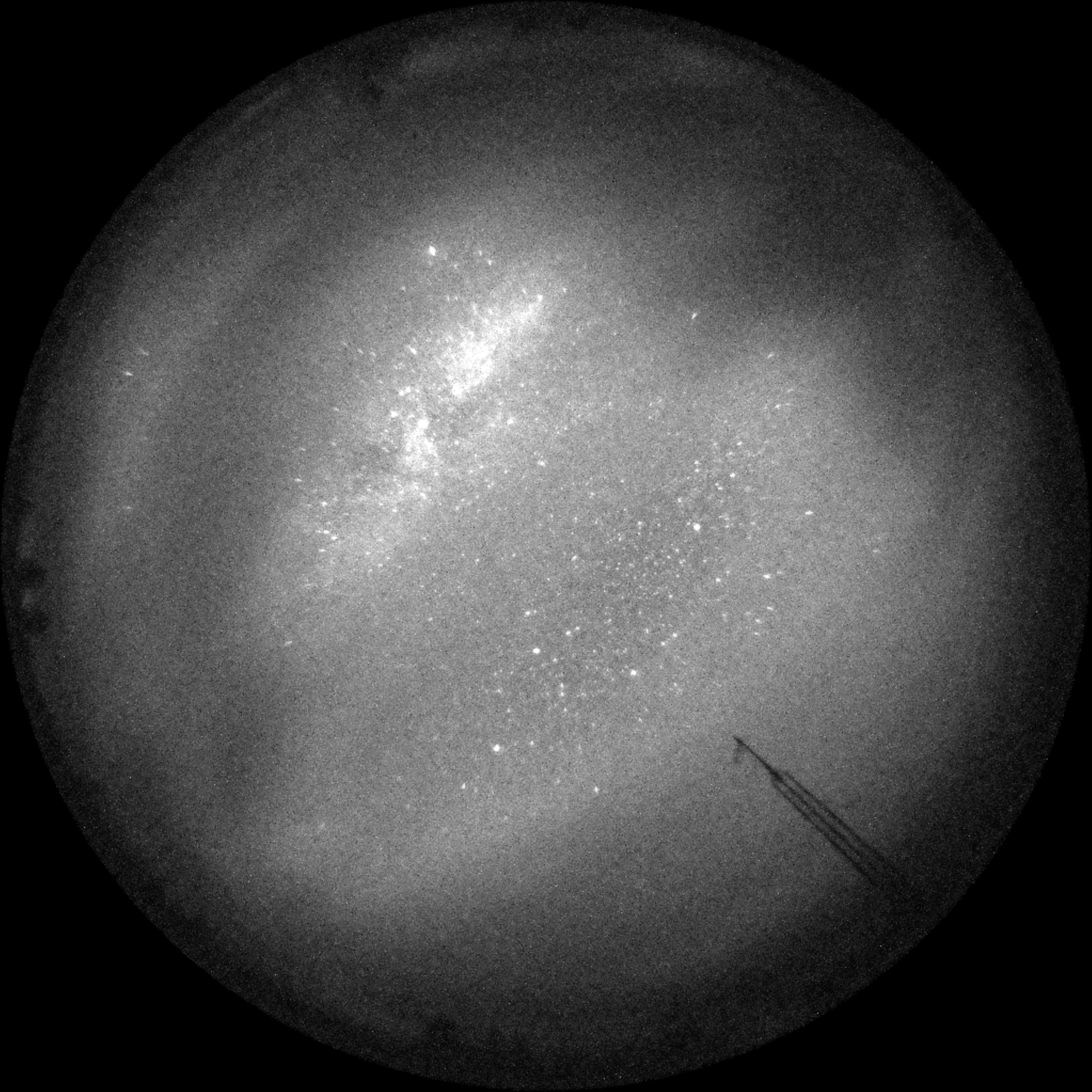}
\caption{Image of partly cloudy night sky from San Pedro Martir. Moon and clouds are visible close to horizon.}
\label{nightSkyClouds}
\end{center}
\end{figure}

The ASC system takes full sky images each $5$ minutes (resp. $10$ minutes - Aar and SAC). Examples of clear and partly cloudy night sky are shown in Figure \ref{nightSky} and \ref{nightSkyClouds}. The on-site analysis (raw cloudiness analysis) follows immediately, the result is saved to two files. The first file consists of a compressed image and online analysis results, the second one consists of a RAW image and relevant data (UTC time, temperature of CCD, Moon and Sun position). 

\begin{figure}[t]
\begin{center}
\includegraphics[width=0.45\textwidth]{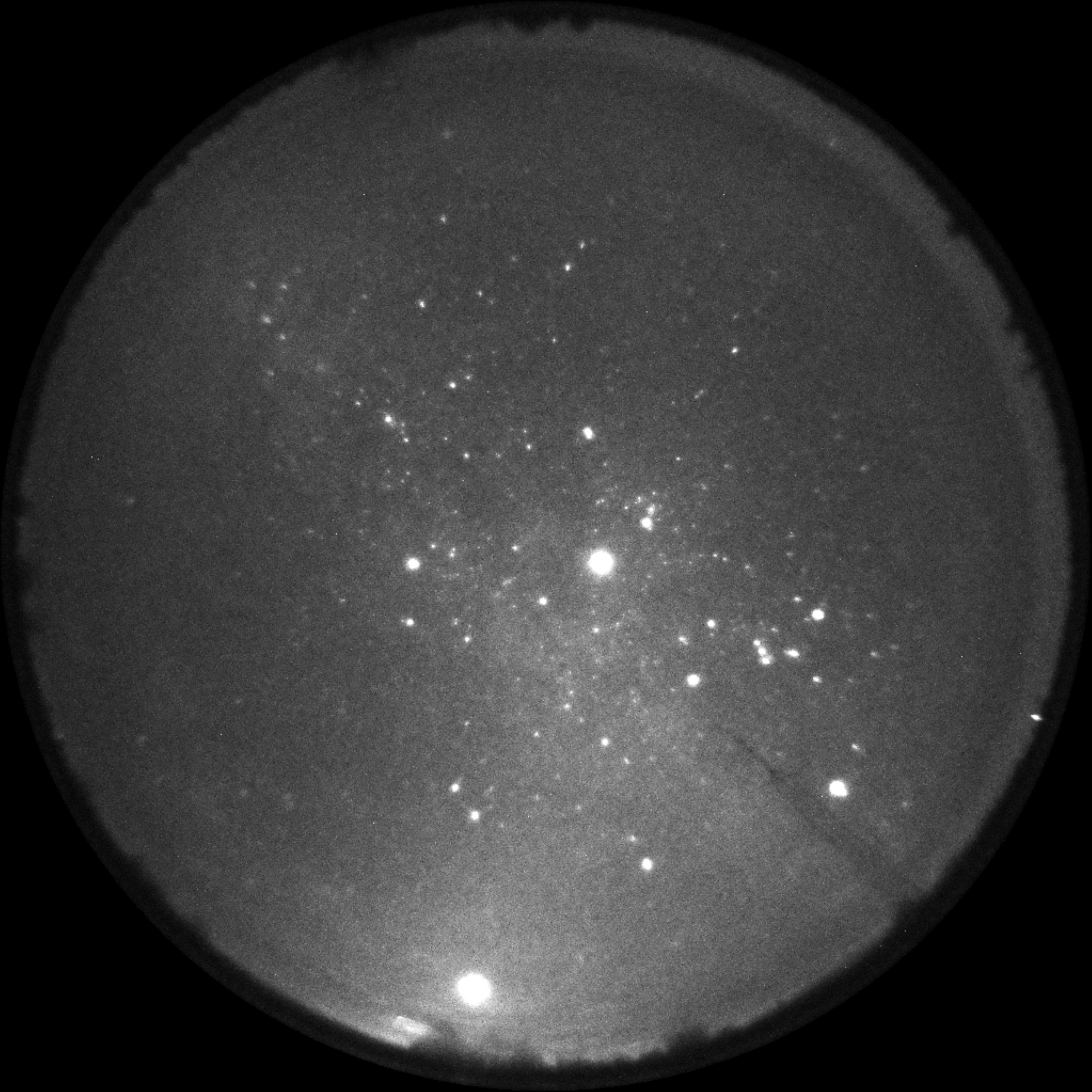}
\caption{Fog or haze affects the visibility of night sky, the nearby tower is not visible (the visibility fades with the fog/haze).}
\label{fog}
\end{center}
\end{figure}

\begin{figure}[!b]
\begin{center}
\includegraphics[width=0.45\textwidth]{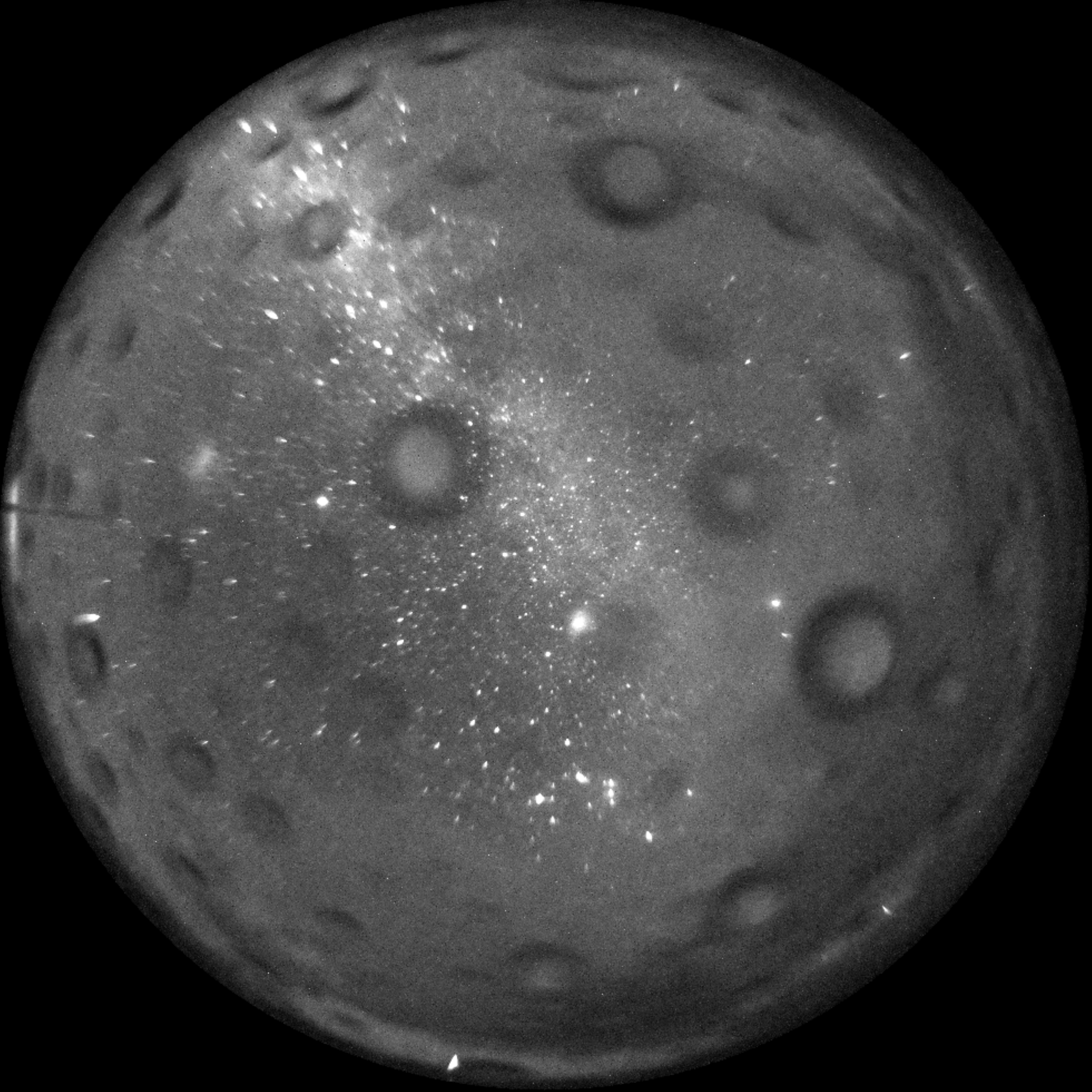}
\caption{Lens of the ASC covered with water drops.}
\label{water}
\end{center}
\end{figure}

\begin{figure*}[!t]
  \centering
  \includegraphics[width=\textwidth]{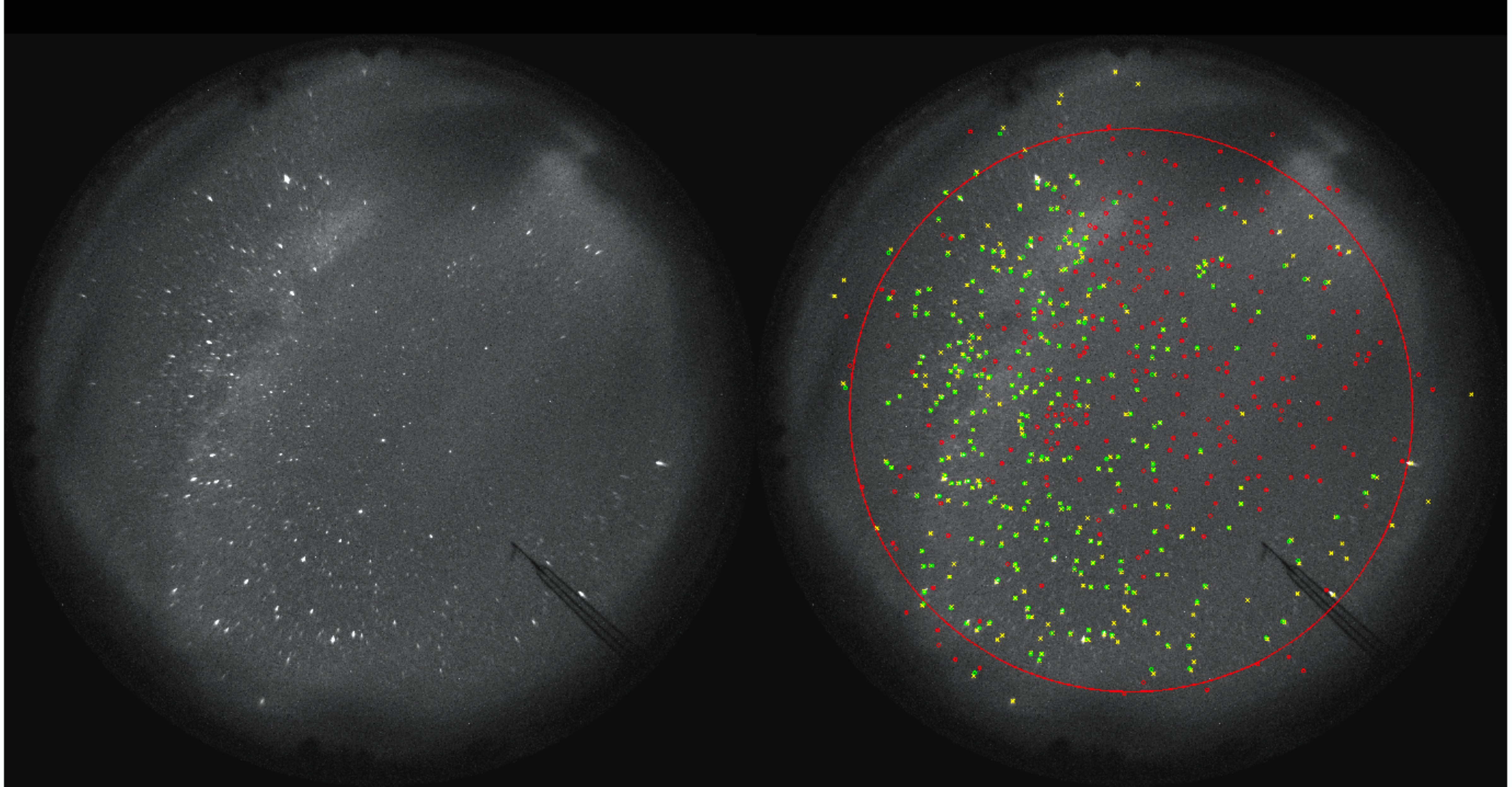}
  \caption{An example of analysis of partly cloudy night sky. The image on the left shows RAW image of the sky. The image on the right shows the analysis results, the yellow crosses indicate the detected stars, green circles catalogue stars and red circles represent catalogue stars without detected pairs - the region covered by cloudiness. The {\color{red} RED} circle indicates the limiting zenith angle ($60^{\circ}$).   }
  \label{cloudiness_analysis}
 \end{figure*} 

The measurement process consists of the following steps. 
Initially, every exposure of a night sky is followed by a dark frame
image so that subtraction of one from the other enables significant
reduction of noise in the image. The horizontal coordinates of each 
image pixel are then calculated with the calibration data and calibration
constants used to determine the horizontal coordinates, azimuth and elevation 
of each image pixel. The total charge of each pixel is proportional to the light
intensity, but the intensity of the final image varies with
respect to zenith angle due to vignetting and distortion of the
lens. The calibration of this zenith angle intensity dependence
is obtained by considering extinction in air masses, optical distortions and 
vignetting variation as a function of zenith angle. The resulting attenuation at, for
example, $60$ degrees zenith angle corresponds to $2$ mag.

To estimate the cloud fraction, we investigate the presence of stars in the field of view of the camera. We separate the full sky to approximately 70 segments and analyze each segment of the sky individually to obtain the cloudiness. As a star catalogue we use Yale Bright Star Catalog BSC$5$ \cite{bib:starcat}, particularly the visual magnitues of the stars. First we estimate the position of each segment on the sky using the data from the calibration. This information is then further improved using the brightest stars from the catalogue in each segement. Once the position of a segment on the sky is fixed, the stars for the segment are read from the catalogue and selected according to the expected sensitivty at the given zenith angle. For every star in the catalogue, we look for a detected star within the angular limit of $1$ deg. If a star that fits this criteria is found, then the catalogue and detected stars are flagged as paired. Figure \ref{cloudiness_analysis} shows a result from cloudiness analysis of partly cloudy night sky. The ratio of paired / unpaired stars (for all segments) gives final cloudiness of the night sky. Typically, we check about 900 catalogue stars for the whole observed sky (up to $60^{\circ}$ which is the maximal operating zenith angle of Cherenkov telescopes). The error of the algorithm was calculated using artificial cloud simulations and the total error of the cloudiness calculation is $\pm 5 \%$.  The cloudiness of the night sky is computed up to the limiting angle with $10 ^{\circ}$ steps.     

Brightness of the night sky is evaluated directly from a light flux. The system has been calibrated using a darkness sky measurement tool Unihedron SQM-LE \cite{bib:sqm}. The night sky was scanned overnight and recorded data both from the camera and SQM instrument were compared to each other. Subsequently the conversion formula is obtained. The data represent the brightness of the night sky in $mag/arcsec^2$ and in visible light spectrum.

\subsection{Limits of the cloudiness analysis}

\begin{figure}[t]
\begin{center}
\includegraphics[width=0.45\textwidth]{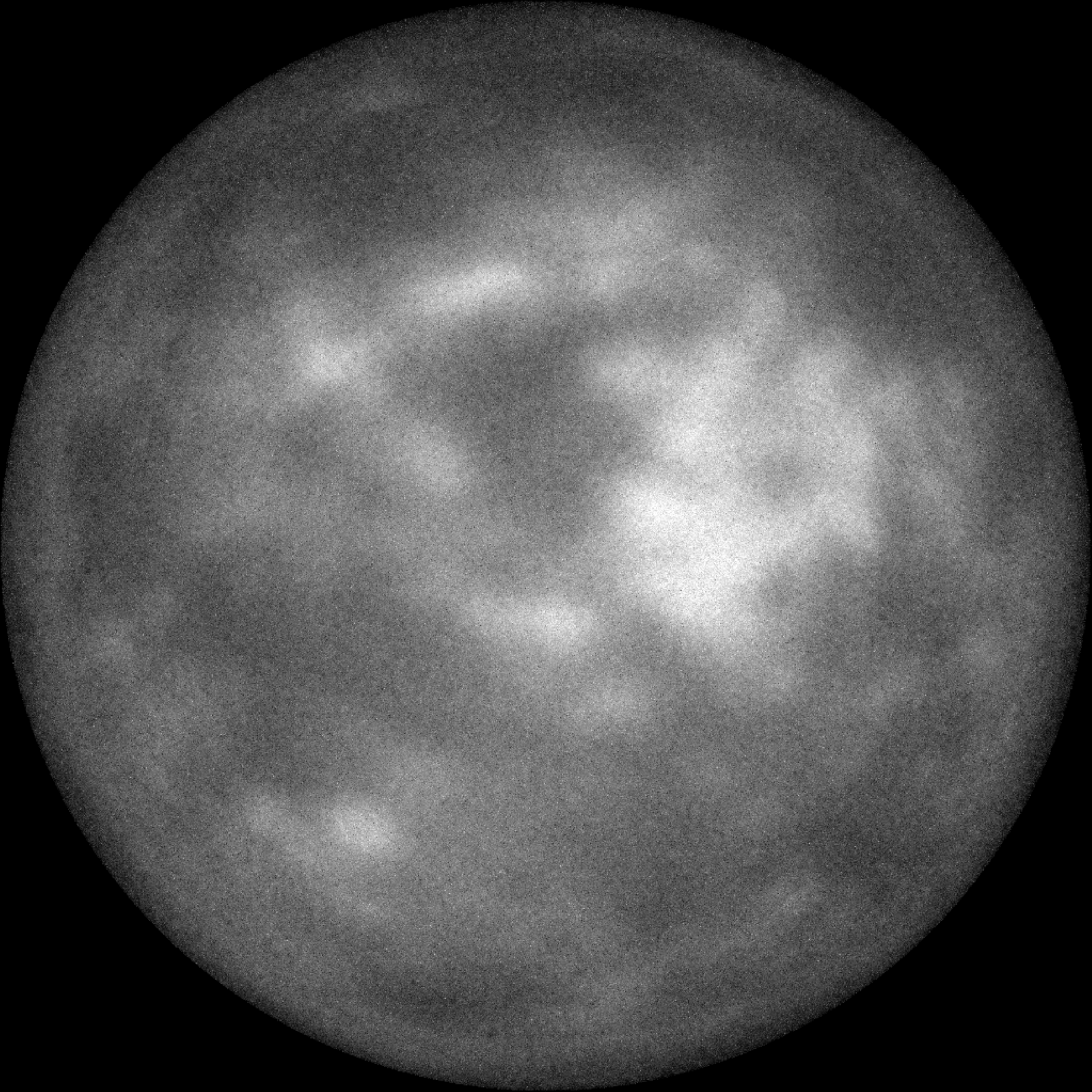}
\caption{Snow covers the whole FOV of the ASC and no star is visible in this case.}
\label{snow}
\end{center}
\end{figure}

The measurement using the ASC is limited by local weather conditions, camera misting and Moon position. The local weather limitations mean for example rain or snow, fog, haze and water condensation on the camera. Examples of rain, fog and snow are shown in Figures \ref {water}, \ref{fog} and \ref{snow} respectively.
The algorithm usually fails if the lens of the ASC is obscured in some way,
for example by snow, water resulting from rain fall, haze or fog
surrounding the Atmoscope, or even bird droppings from birds 
found locally at the Aar and Tenerife sites. The presence of the Moon saturates the image and increases the background in the image, and consequently the algorithm for cloud detection could not be used at such times.

\section{Applications, operation and advantages}

The main advantage of the all-sky camera is that it is a fully autonomous and self-reliant tool which can be used to monitor the night sky in even the world's most remote places, which happen to be just the ones most promising for astronomically related applications. When using solar power, the only requirements for the deployment of the system are the ability to transport the device and the solar power system to the location. The availability of internet connection is useful in order to ensure fast response to possible issues, but is by no means a requirement. 
The autonomy of the camera significantly contributes to the extremely low cost for its operation. The costs involved are concentrated mainly in the deployment phase, namely the costs of the parts and assembly of the camera, shipping (varies depending on destination, the shippable weight being $10$ kg), the solar power system, if there is no pre-existing power on-site (the $180$ W required by the system) and installation (two people working for two or three days). After installation, the camera can operate without maintenance for years. However, it is advantageous to have local support for the case of unexpected external hindrances, but in most cases, a non-specialist local personnel is sufficient to fix most of the possible issues.

\section{Conclusions}

ASC camera for night sky measurement was presented as a versatile remote system for cloudiness and night sky brightness characterization. Eight ASC's are being used to measure night sky parameters at the candidate sites of the Cherenkov Telescope Array (CTA) gamma-ray observatory. The data will be used to characterize the candidate site and to use the data in the selection process for Southern and Northern CTA site.

\vspace*{0.5cm}
\footnotesize{{\bf Acknowledgment:}{ This work is supported by Ministry of Education of the Czech Republic, under the projects EUPROII LE13012 and Mobility 7AMB12AR013. We gratefully acknowledge support from the agencies and organizations
listed in this page: http://www.cta-observatory.org/?q=node/22.}}

\end{document}